\documentclass[12pt]{article}

\usepackage[latin1]{inputenc}
\usepackage[T1]{fontenc}
\usepackage{a4}
\usepackage{amsmath}  
\usepackage{amssymb}
\usepackage{color}
\usepackage{bbm}
\usepackage[a4paper,colorlinks,citecolor=blue,linkcolor=blue,urlcolor=blue,pdfpagemode=None]{hyperref}  

  \vfuzz \hfuzz
\newcommand{\D}{\text{d}}
\newcommand{\E}{\text{e}}
\newcommand{\I}{\text{i}}
\newcommand{\C}{\mathbbm{C}}
\newcommand{\N}{\mathbbm{N}}

\newcommand{\R}{\mathbbm{R}}
\newcommand{\Z}{\mathbbm{Z}}

\newcommand{\be}{\begin{equation}}
\newcommand{\ee}{\end{equation}}
\newcommand{\NO}[1]{\boldsymbol{:} #1 \boldsymbol{:}}
\newcommand{\OPE}{operator product expansion}
\newcommand{\VOA}{vertex operator algebra}

\allowdisplaybreaks

\numberwithin{equation}{section}

\title{{\bf Nonmeromorphic operator product expansion and $\boldsymbol{C_2}$-cofiniteness for a family of $\boldsymbol{\mathcal{W}}$-algebras}}
\author{Nils Carqueville\thanks{{\tt \href{mailto:nils@th.physik.uni-bonn.de}{nils@th.physik.uni-bonn.de}}}\quad\ and\quad Michael Flohr\thanks{{\tt \href{mailto:flohr@th.physik.uni-bonn.de}{flohr@th.physik.uni-bonn.de}}}
\\[0.5cm]
  {\normalsize\slshape Physikalisches Institut}\\[-0.1cm]
  {\normalsize\slshape University of Bonn}\\[-0.1cm]
  {\normalsize\slshape Nussallee 12}\\[-0.1cm]
  {\normalsize\slshape 53115 Bonn, Germany}}
\date{}

\begin{document}

\maketitle

\begin{abstract}
We prove the existence and associativity of the nonmeromorphic \OPE\ for an infinite family of \VOA s, the triplet $\mathcal{W}$-algebras, using results from $P(z)$-tensor product theory. While doing this, we also show that all these \VOA s are $C_2$-cofinite. 
\end{abstract}

\section{Introduction}

The notion of \OPE\ is fundamental in (quantum) field theory; this notion was originally introduced by Wilson \cite{wil69} and Kadanoff \cite{kad69}. Physically, it describes the short distance behaviour of the product of two quantum fields $\Phi_1(z_1)$ and $\Phi_2(z_2)$ when it is evaluated near the singularity at $z_1 = z_2$ and therefore encodes part of the local structure of the theory. The singularity arises because the fields are distributions. 

In particular, products of fields occur in correlation functions, which eventually allow to compute observables which can then be compared with experimental data. One important advantage of \OPE, viewed as a tool that expands the product of \textit{two} fields into a series in which each summand involves only \textit{one} field, is that in this way $n$-point functions can be expressed in terms of $(n-1)$-point functions. This does not only tremendously facilitate concrete computations, but it also structures the theory conceptually. 

While for generic spacetime dimensions a rigorous formulation and treatment of \OPE\ is very difficult, in two-dimensional conformal field theory the situation is much clearer (which is one of many reasons to study conformal field theory, apart form its prominent role and successful application in string theory and statistical physics). 

A natural framework to deal with two-dimensional conformal field theory is that of \VOA s \cite{fhl}, \cite{LeLi} and related concepts. In this language, the \OPE\ of fields operating on a \textit{fixed} module (or representation) $W$ for a \VOA\ $V$ is expressed in the well-known result
\be\label{meroOPE}
\iota_{12}^{-1} \left\langle w' , Y(u,x_1) Y(v,x_2) w \right\rangle = \left( \iota_{20}^{-1} \left\langle w' , Y\left( Y(u,x_0)v,x_2\right) w \right\rangle \right) \Big|_{x_0=x_1-x_2} \; ,
\ee
where $u,v\in V$, $w\in W$, $w'\in W' = \coprod_{n\in\Z}(W_{[n]})^{*}$, $\langle\,\cdot\,,\,\cdot\,\rangle$ denotes the natural pairing between $W'$ and $W$, and $\iota_{ij}$ denotes the operation of expanding a function of $x_i$ and $x_j$ such that only finitely many negative powers of $x_j$ appear in the expansion. In the physics literature, (\ref{meroOPE}) usually is given in disguise as
$$
Y(u,z_1) Y(v,z_2) \sim \sum_{n=0}^N Y(u_n v,z_2) (z_1-z_2)^{-n-1} \; ,
$$
with some $N\in\N$ depending on $u$ and $v$, where the iteration of two vertex operators has been expanded, only the terms singular in $(z_1-z_2)$ are given, and this relation should be implicitly understood to be used inside a correlation function. 

The \OPE\ (\ref{meroOPE}) is referred to as meromorphic since vertex operators acting on modules only involve integer powers of formal variables. But much of the interesting information in both physical and mathematical theories is hidden in the way different modules ``interact''. So instead of vertex operators that always act from the underlying \VOA\ on a given module, \textit{intertwining operators} have to be considered, which mediate between three (possibly distinct) modules and do not necessarily depend meromorphically on their variables. 

While the existence and associativity of the \OPE\ for vertex operators $Y(\,\cdot\,,x)\,\cdot$ follows rather easily from the axioms, this is not the case for intertwining operators $\mathcal{Y}(\,\cdot\,,x)\,\cdot$. In fact, it is a deep result of the intricate $P(z)$-tensor product theory of Huang and Lepowsky developed in \cite{hl9309a}--\cite{hl9505} and \cite{hua9505a}. In \cite{hlz}, Huang, Lepowsky and Zhang generalized their results such that they could drop the condition of rationality and of semisimplicity of the action of $L_0$. In this way, a logarithmic dependence on the variables may appear in intertwining operators and correlation functions. This also seems to be the most successful rigorous treatment of logarithmic conformal field theory (e.g., see the review articles \cite{gab0111} and \cite{flohr0111}) so far. 

\vspace{0.2cm}

Now we recall some of the results of \cite{hlz} which are central for this work. For any generalized module $W$ of a \VOA\ $V$, its restricted dual $W' = \coprod_{n\in\Z}(W_{[n]})^{*}$ can be endowed with a module structure, where the associated vertex operator $Y'(\,\cdot\,,x)\;\cdot$ is defined by
$$
\left\langle \vphantom{\left(-x^{-2}\right)^{L_0}} Y'(v,x)w' , w \right\rangle = \left\langle w' , Y\left( \E^{x L_1} \left(-x^{-2}\right)^{L_0} v , x^{-1} \right) w \right\rangle
$$
for all $v\in V$, $w\in W$ and $w'\in W'$. The pair $(W',Y')$ is called the contragredient module to $(W,Y)$, and the contravariant functor $(\,\cdot\,)': (W,Y) \mapsto (W',Y')$ is called the contragredient functor. (Here and in the following, we understand a generalized module in the way a module is usually defined as for example in \cite{LeLi}, except that the action of $L_0$ may have a nilpotent part, i.e.~the homogeneous subspaces $W_{[n]} = \{w\in W \,|\, (L_0-n)^m w = 0 \text{ for } m \gg 0\}$ are generalized eigenspaces which are in particular assumed to be finite-dimensional. Sometimes we call such a structure simply a module, omitting the attribute generalized.) Also, let $\mathcal{C}$ denote an as yet unspecified full subcategory of the category whose objects are all (generalized) modules of a given \VOA\ $V$, that is closed under the contragredient functor. 

In \cite{hlz}, the authors use the $P(z)$-tensor product theory and carefully establish several conditions for the existence and associativity of the \OPE\ for logarithmic intertwining operators. Instead of repeating all the steps in their argument here, we refer to their paper for details and give only a summary of the results concerning the \OPE. Indeed, once an adequate subcategory $\mathcal{C}$ has been chosen, it follows from Proposition 5.13 and Theorems 6.1 to 6.3 in \cite{hlz} that if its objects satisfy the following three conditions, then the nonmeromorphic \OPE\ exists and is associative. These conditions are: 
\begin{enumerate}
\item\label{nummer1} All generalized $V$-modules in $\text{ob}\,\mathcal{C}$ are $C_1$-cofinite, i.e.~for all $W\in \text{ob}\,\mathcal{C}$ the space $W/C_1(W)$ is finite-dimensional with
\[
C_1(W) = \text{span}\left\{ u_{-1}w \; \Big| \; u\in \coprod_{n>0} V_{(n)} , \, w\in W \right\} \; . 
\]
\item\label{nummer2} All generalized $V$-modules in $\text{ob}\,\mathcal{C}$ are quasi-finite-dimensional, i.e.~for all  $W\in \text{ob}\,\mathcal{C}$, $$ \text{dim}  \coprod_{n<N} W_{[n]} < \infty \quad \text{ for all } N\in\R \; . $$
\item\label{nummer3} Every object which is a finitely generated lower-truncated generalized $V$-module, except that it may have infinite-dimensional homogeneous subspaces, is an object in $\mathcal{C}$. 
\end{enumerate}
Note that the basic structure in \cite{hlz} are conformal vertex algebras, which are the same as vertex operator algebras except that they might not satisfy the lower-truncation condition and their homogeneous subspaces may be infinite-dimensional, and the same is true for their (generalized) modules. This is why the objects may not be assumed to have finite-dimensional homogeneous subspaces in condition (iii), but this must be proven. 

The precise statement of the assertion is that under the above conditions, for any $W_1, W_2, W_3, W_4'$ and $M$ in $\text{ob}\,\mathcal{C}$, any $P(z_1)$-intertwining map $I_1$ of type $\binom{W_4}{W_1\;\, M}$ and any $P(z_2)$-intertwining map $I_2$ of type $\binom{M}{W_2\;\, W_3}$, there is a $P(z_2)$-intertwining map $I$ of type $\binom{W_4}{W_1\boxtimes_{P(z_0)} W_2\;\,W_3}$ such that
\be\label{nonmeroOPE}
\left\langle w_4' , I_1 (w_1,z_1) I_2(w_2,z_2) w_3 \right\rangle = \left\langle w_4' , I(w_1 {\scriptstyle \boxtimes}_{P(z_0)} w_2 , z_2) w_3 \right\rangle 
\ee
for all $w_1\in W_1$, $w_2\in W_2$, $w_3\in W_3$ and $w_4'\in W_4'$. 

The formal similarity between (\ref{meroOPE}) and (\ref{nonmeroOPE}) is obvious. The subtle point is that Huang, Lepowsky and Zhang really proved that in the \OPE\ of two logarithmic intertwining maps, there only appear powers of the variables and their logarithms (with no further dependence on the variables), while in the physics literature this is usually assumed without proof. With this information one can then try to find differential equations that are solved by the matrix elements of the product of two logarithmic intertwining maps at, say, $z_1$ and $z_2$. (Such differential equations always do exist by a theorem of Huang \cite{hua0206}, \cite{hlz}.) Expanding the solution in $z_0 = z_1 - z_2$ and $z_2$, one arrives at the desired \OPE, evaluated inside a matrix element -- and only this way is the \OPE\ well-defined. 

\vspace{0.2cm}

In the present paper, we show that a certain family $\{\mathcal{W}(2,(2p-1)^{\times 3})\}_{p\geq 2}$ of $\mathcal{W}$-algebras satisfies the above conditions. The \VOA\ $\mathcal{W}(2,3^{\times 3})$ is historically the first example of a ``rational'' \textit{logarithmic} conformal field theory in the sense of \cite{gk9606}, i.e.~a finite set of its modules closes under fusion. From the explicitly known characters for the \VOA s $\mathcal{W}(2,(2p-1)^{\times 3})$ it follows that they all display logarithmic features, see \cite{flohr9509}. This is one of the motivations for us to study these $\mathcal{W}$-algebras in more detail here. 
 
The main work of our proof is to establish that each member of this family is $C_2$-cofinite, and from this fact condition (i) follows as we will show (the other two conditions are easy to check). Thus, we do not only prove the existence and associativity of the \OPE\ as described above for each $\mathcal{W}(2,(2p-1)^{\times 3})$, but we also establish one of the most useful and interesting properties in the study of \VOA s for these $\mathcal{W}$-algebras. 

Indeed, the condition of $C_2$-cofiniteness was introduced by Zhu in \cite{zhu96} and subsequently used to prove the convergence and modular invariance of characters for certain \VOA s, and it is also related to his famous associative algebra $A(V)$. But $C_2$-cofiniteness is also important because of its close relation to rationality and regularity. It was proven by Li in \cite{li9807} that any regular \VOA\ is also $C_2$-cofinite, and Abe, Buhl and Dong were able to show in \cite{abd0204} that regularity is equivalent to rationality (in the sense of complete reducibility of all modules) and $C_2$-cofiniteness together for \VOA s of the form $V=\coprod_{m\in\N}V_{(m)}$ with $V_{(0)}=\C\Omega$. Even a conjecture was formulated that rationality (in the sense of complete reducibility of all modules), regularity and $C_2$-cofiniteness are equivalent properties for \VOA s, but $\mathcal{W}(2,3^{\times 3})$ actually serves as a counter-example. This was known from the work of Kausch, and Abe explicitly noted it in \cite{abe0503}. Nevertheless, one may still conjecture the equivalence of $C_2$-cofiniteness and ``rationality'' in the sense of \cite{gk9606}, i.e.~a finite set of modules closes under fusion. 

Our proof that all triplet algebras $V_{2p-1}=\mathcal{W}(2,(2p-1)^{\times 3})$ are $C_2$-cofinite adds credibility to this conjecture. Indeed, from the $C_2$-cofiniteness of $V_{2p-1}$ it follows that the Zhu algebra $A(V_{2p-1})$ is finite-dimensional, and because of this there are only finitely many equivalence classes of indecomposable $A(V_{2p-1})$-modules. This together with the strong restrictions coming from the structure of $\mathcal{W}$-algebras suggests that the assumedly equivalent properties both hold for all triplet algebras. 

What makes this relationship particularly interesting is the fact that while rationality explicitly concerns the modules for a \VOA, the $C_2$-cofiniteness condition can be studied solely in terms of the \VOA\ itself, without reference to any modules. 

\vspace{0.2cm}

The remaining structure of the present paper is as follows. In section \ref{Preliminaries} we briefly recall a few basic properties concerning $C_n$-cofiniteness; we also explain what we mean by a $\mathcal{W}$-algebra in general and give some useful results. For standard notions concerning \VOA s and for details about concepts like (logarithmic) intertwining maps (which are in one-to-one correspondence to (logarithmic) intertwining operators) or the $P(z)$-tensor product, we refer the reader to the above-mentioned literature. Then, in section \ref{mainresults}, we first prove our result for the intimately known \VOA\ $\mathcal{W}(2,3^{\times 3})$ and then show how the respective arguments can be generalized to all members of the family $\{\mathcal{W}(2,(2p-1)^{\times 3})\}_{p\geq 2}$. 

\vspace{0.2cm}

\textit{Acknowledgments}: Nils Carqueville thanks Yi-Zhi Huang for explaining several aspects of $P(z)$-tensor product theory to him. He also thanks Julia {Voelskow} for taking interest in this work. Both authors thank James Lepowsky and Geoffrey Buhl for many valuable comments. The research of Michael Flohr is supported by the European Union Network
HPRN-CT-2002-00325 (EUCLID).

\section{Preliminaries}\label{Preliminaries}

\subsection*{The subspaces $\boldsymbol{C_n(W)}$}

In order to prove that the above conditions are satisfied for a given \VOA\ $V$ and appropriately chosen $\mathcal{C}$, we first give two general properties of the spaces $C_n(W)$, where $W$ is any generalized $V$-module. For $n\geq 2$, $C_n(W)$ is defined as $C_n(W) = \text{span}\{ u_{-n}w \,|\, u\in V , \, w\in W \}$, and for $n=1$, we have $C_1(W) = \text{span}\{ u_{-1}w \,|\, u\in \coprod_{n>0} V_{(n)} , \, w\in W \}$. The space $W$ is called $C_n$-cofinite if $\text{dim}  (W/C_n(W))<\infty$. Thus, because of the $L_{-1}$-derivative property
$$
Y\left( L_{-1}^m v,x\right) = \frac{\D^m}{\D x^m} Y(v,x) \; , 
$$
it directly follows by comparing coefficients that
\be\label{CminCn}
v_{-m-1} = \frac{1}{m} \left( L_{-1} v \right)_{-m} = \frac{1}{m!} \left(L_{-1}^m v \right)_{-1} \quad \text{ for all } m\in\Z_+ \; , 
\ee
and hence every $C_m$-cofinite generalized  $V$-module $W$ is also $C_n$-cofinite for all $m\geq n\geq 1$. For $n=1$, this can also be expressed by writing 
\be\label{ausmmach1}
C_1(W) = \text{span}\left\{ u_{-m}w \; \Big| \; u\in \coprod_{n>0} V_{(n)} , \, w\in W , \, m\in\Z_+ \right\} \; .
\ee

Secondly, $C_n(W)$ is invariant under the action of $v_m$ for all $v\in V$ and $m\leq 0$. To prove this, one only needs to look at the well-known commutation relation
\be\label{modenkommie}
v_m u_{-n} w = u_{-n} v_m w + \sum_{i\in\N} \binom{m}{i} \left(v_i u\right)_{m-n-i} w
\ee
which follows from the Jacobi identity by performing the usual residue operation. For $m\leq 0$ the right-hand side of (\ref{modenkommie}) obviously is an element of $C_n(W)$ because of the relation (\ref{CminCn}), and so this must also be true for the left-hand side. The result is
\be\label{c1invariant}
v_m C_n(W) \subset C_n(W) \quad \text{ for all } v \in V \, , \; m\in\Z_{\leq 0} \, , \; n\in\Z_+ \; .
\ee

Finally, another useful relation that follows from the Jacobi identity is
\be\label{modeniter}
(u_m v)_n = \sum_{i\in\N} (-1)^i \binom{m}{i} u_{m-i} v_{n+i} - \sum_{i\in\N} (-1)^{i+m} \binom{m}{i} v_{m+n-i} u_i \; . 
\ee

\subsection*{$\boldsymbol{\mathcal{W}}$-algebras}

A $\mathcal{W}$-algebra of type $\mathcal{W}(2,h_1,\ldots,h_m)$ is a \VOA\ which has a generating set consisting of the vacuum $\Omega$, the conformal vector $\omega$ of weight 2 and $m$ additional primary vectors $W^i$ of weight $h_i$, $i\in\{1,\ldots,m\}$. The vertex operators or fields associated to these vectors are simple in the sense that they are not normal-ordered products of other fields. Sometimes the term $\mathcal{W}$-algebra is also used to refer to the algebra of modes instead of the \VOA. 

What will be of particular importance for us is the notion of quasi-primary normal-ordered products which is due to Nahm \cite{nahmx}. When working with $\mathcal{W}$-algebras, we will mainly adopt his notation. For a more detailed exposition, see for example \cite{flohr91}. Here, only those relations are given that are needed for our present purpose. 

The usual normal-ordered product $\NO{\phi_i(x)\phi_j(y)}\, =\phi_i(x)_+\phi_j(y) + \phi_j(y)\phi_i(x)_-$ of two quasi-primary fields $\phi_i$ and $\phi_j$ is not necessarily quasi-primary for $x=y$. One of Nahm's results is that it is always possible to add certain correction terms, yielding a quasi-primary normal-ordered product denoted by $\mathcal{N}(\,\cdot\,,\,\cdot\,)$: 
\begin{align}\label{qpNOP}
\mathcal{N}\left(\phi_j , \partial^n \phi_i \right) & = \sum_{r=0}^n (-1)^r \binom{n}{r} \binom{2(h_i + h_j + n - 1)}{r}^{-1} \binom{2 h_i + n - 1}{r} \nonumber \\
& \qquad \cdot \partial^r N^{(h_i + n + r)}\left( \phi_j , \partial^{n-r} \phi_i \right) \nonumber \\
& \qquad - (-1)^n \sum_{\{k \,|\, h(ijk) \geq 1\}} C_{ij}^k \binom{h(ijk) + n - 1}{n} \nonumber \\
& \qquad \cdot \binom{2(h_i + h_j + n - 1)}{n}^{-1} \binom{2 h_i + n - 1}{h(ijk) + n} \binom{\sigma(ijk) - 1}{h(ijk) - 1}^{-1} \nonumber \\
& \qquad \cdot \frac{\partial^{h(ijk) + n} \phi_k}{(\sigma(ijk) + n)(h(ijk) - 1)} \; .
\end{align}
Here, $\{\phi_k\}_k$ is the family of quasi-primary fields of the corresponding $\mathcal{W}$-algebra, $h_k$ are their respective weights, $h(ijk):=h_i+h_j-h_k$ and $\sigma(ijk):=h_i+h_j+h_k-1$. The structure constants $C_{ij}^k$ are defined such that $\sum_l C_{ij}^l d_{lk} = C_{ijk}$ with
$$
C_{ijk} = \left\langle \Omega' , (\phi_k)_{+h_k} (\phi_i)_{-h_k+h_j} (\phi_j)_{-h_j} \Omega \right\rangle \quad \text{ and } \quad d_{ij} = \left\langle \Omega' , (\phi_i)_{+h_i} (\phi_j)_{-h_j} \Omega \right\rangle \; , 
$$
and the $N^{(\,\cdot\,)}$-product is defined by the relations
\begin{align}\label{NOPModen}
N^{(m)}(\phi,\psi)(x) & = \sum_{n\in\Z} x^{-n-h_\phi-h_\psi} N^{(m)}(\phi,\psi)_n \; , \nonumber \\
N^{(m)}(\phi,\psi)_n & = \sum_{k<m} \phi_{n+k} \psi_{-k} + \sum_{k\geq m} \psi_{-k} \phi_{n+k}
\end{align}
for any $m\in\Z$. The quasi-primary normal-ordered product of more than two fields is defined recursively, for example $\mathcal{N}(\phi_i,\phi_j,\phi_k) = \mathcal{N}(\phi_i,\mathcal{N}(\phi_j,\phi_k))$. If the product of a field with itself is considered the notation is simplified, for example $\mathcal{N}(\psi,\psi) = \mathcal{N}(\psi^2)$. 

Furthermore, in this notation the commutators of modes are given by
\be\label{WKommutator}
\big[ (\phi_i)_m , (\phi_j)_n \big] = d_{ij} \delta_{m+n,0} \binom{h_i + m - 1}{2 h_i - 1} + \sum_{\{k \,|\, h(ijk) \geq 1\}} C_{ij}^k \, p_{h_i,h_j,h_k}(m,n) (\phi_k)_{m+n}
\ee
in terms of the polynomials
$$
p_{h_i,h_j,h_k}(m,n) = \sum_{r,s\in\N} \delta_{r+s,h(ijk)-1} a^r_{ijk} \binom{m + n - h_k}{r} \binom{h_i - n - 1}{s} 
$$
with
$$
a^r_{ijk} = \binom{2 h_k +r - 1}{r}^{-1} \binom{h_i + h_k - h_j + r - 1}{r} \; . 
$$

In the next section, also the formal power series known as the character
$$
\chi_V(q) = \text{tr}_V q^{L_0 - c/24} = q^{-c/24} \sum_{n\in\N} \text{dim}  V_{(n)} \, q^n
$$
of the \VOA\ $V=\mathcal{W}(2,h_1,\ldots,h_m)$ will be useful. In our proof of the existence of certain singular vectors, we will compare this character with the character of the vacuum Verma module of the $\mathcal{W}$-algebra. This is the induced module
$$
U(\mathcal{W}(2,h_1,\ldots,h_m)) \otimes_{U(\mathcal{W}(2,h_1,\ldots,h_m)_{(+)})} \C_c \; ,
$$
where $U(\,\cdot\,)$ denotes the universal enveloping algebra of the $\mathcal{W}$-algebra, the space $\mathcal{W}(2,h_1,\ldots,h_m)_{(+)}$ is defined by 
$$
\mathcal{W}(2,h_1,\ldots,h_m)_{(+)} = \coprod_{n\leq 1} \C L_{-n} \oplus \coprod_{i=1}^m \coprod_{n_i\leq h_i-1} \C W_{-n_i}^i \; ,
$$
and $\C_c$ is the trivial $\mathcal{W}(2,h_1,\ldots,h_m)_{(+)}$-module of central charge $c$. In other words, the vacuum Verma module is generated freely by the action of the modes $L_n$ and $W_n^i$ on a nonzero element $\Omega$ in $\C_c$, subject to the restrictions
\be\label{Vermarestrictions}
L_n \Omega = 0 \text{ for all } n \geq - 1 \quad \text{and} \quad W_n^i \Omega = 0 \text{ for all } n \geq - h_i + 1 \; . 
\ee
Because of these restrictions, the dimensions of the homogeneous subspaces $V_{(n)}$ are smaller than $p(n)$, where $p(n)$ is the number of partitions of $n$ into sums of positive integers, generated by the function 
$$
(\varphi(q))^{-1} = \prod_{n\geq 1} \left(1-q^n\right)^{-1} = \sum_{n\in\N} p(n) q^n \; .
$$
Taking the restrictions (\ref{Vermarestrictions}) into account, the vacuum Verma module character is given by
\be\label{vacVermachar}
\chi_V^{\text{{\tiny Verma}}}(q) = q^{-c/24} (\varphi_2(q))^{-1} \prod_{i=1}^m \big( \varphi_{h_i}(q) \big)^{-1} \; ,
\ee
where we have introduced the generating functions $\varphi_k$, $k\geq 2$, as truncated $\varphi$-functions: 
\be\label{truncvarphi}
\varphi_k (q) = \prod_{n\geq k}\left( 1-q^n \right) = \varphi(q) \prod_{l=1}^{k-1} \left(1-q^l \right)^{-1} \; . 
\ee

\section{Properties of the triplet algebras}\label{mainresults}

\subsection*{The triplet algebra at $\boldsymbol{c=-2}$}

With the general relations of the preceding section at hand, we now choose a \VOA\ $V$ and a suitable category $\mathcal{C}$ for which we will prove the conditions for the existence and associativity of the nonmeromorphic \OPE. We take $V=V_3$ which is the triplet algebra with central charge $c=-2$ \cite{kau91}, \cite{flohr9509}, \cite{gk9606} and $\mathcal{C}$ such that its objects are exactly all finitely generated lower-truncated generalized $V$-modules. (Note that in this context, the notion of such a module of $V$ by definition encompasses a possible Jordan cell structure in the $L_0$-grading.) In particular, this choice includes all (generalized) highest weight modules, but also those on which $L_1$ acts only nilpotently (and not necessarily trivially) on the generating vector.

The triplet algebra at $c=-2$ is a $\mathcal{W}$-algebra of type $\mathcal{W}(2,3^{\times 3})$. It is generated by the modes $L_m$ of the Virasoro field $T(x) = \sum_{m\in\Z} L_m x^{-m-2}$ associated to the vector $\omega$ of weight 2 which implements the conformal symmetry, and the modes $W_m^a$ of a triplet (under the action of the group $SO(3)$) of primary fields of weight 3, $W^a(x) = \sum_{m\in\Z} W_m^a x^{-m-3}$ with $a\in \{\pm 1, 0\}$, which ``maximally extend'' the conformal symmetry, acting on some module $W\in\text{ob}\,\mathcal{C}$. With this notation and by relation (\ref{ausmmach1}), the vectors 
\be\label{LWinC1}
L_{-m+1}w \text{ and } W_{-m}^a w \text{ are in } C_1(W) \text{ for all } w\in W \text{ and all } m\geq 3 \; , 
\ee
while for all other values of $m$, this is not necessarily the case. 

By this choice of $\mathcal{C}$, condition (iii) above is satisfied. The fact that the homogeneous subspaces of the (generalized) modules in $\text{ob}\,\mathcal{C}$ are really finite-dimensional follows from results of Buhl \cite{b0111} on a module spanning set, using the fact that they are finitely generated and all triplet algebras are $C_2$-cofinite, see below. 

Condition (ii) also is satisfied: by the action of any mode $v_n$ with $v\in V$, the weight of an element to which $v_n$ is applied to changes by an integer value, and there are only finitely many vectors that generate $V$, namely $\omega$ and $W^a$ (together with the vacuum $\Omega$). 

In order to see that condition (i) is satisfied as well, i.e.~$C_1(W)$ is finite-codimensional for all $W\in\text{ob}\,\mathcal{C}$, we assume without loss of generality that $W$ is generated by some element $w=w^{(0)}$ together with its (finitely many) ``logarithmic partners'' $w^{(i)}$. Then by (\ref{modenkommie}) and (\ref{modeniter}), every vector in $W$ is a linear combination of elements of the form
\be\label{sternie}
\mathcal{M}_{-m_1} \ldots \mathcal{M}_{-m_k} L_{-1}^M \prod_{a\in\{\pm 1,0\}} \left( \left(W_{-2}^a\right)^{N_2^a} \left(W_{-1}^a\right)^{N_1^a} \left(W_{0}^a\right)^{N_0^a} \right) \mathcal{M}_{n_1} \ldots \mathcal{M}_{n_l} w^{(i)}
\ee
where $\mathcal{M}$ is a placeholder for either $L$ or $W^a$; $M$, $N_0^a$, $N_1^a$, $N_2^a \in \N$, $n_1, \ldots, n_l \in \Z_+$ and $m_1, \ldots, m_k \in \Z_{\geq 2}$ for $\mathcal{M}=L$ while $m_1, \ldots, m_k \in \Z_{\geq 3}$ for $\mathcal{M}=W^a$. 

In the case that $k$ is strictly larger than zero, (\ref{LWinC1}) immediately shows that any such element is in $C_1(W)$. On the other hand, for $k=0$ there are only finitely many possibilities for the term $\mathcal{M}_{n_1} \ldots \mathcal{M}_{n_l} w^{(i)}$ not to vanish because of the lower-truncatedness of $W$. The factor $(W_{0}^a)^{N_0^a}$ can also do no harm as it does not change the generalized weight of the element it is applied to, and each Jordan cell is finite-dimensional by the definition of $\mathcal{C}$, given the fact that condition (ii) is satisfied. 

So what deserves special attention are the powers of $L_{-1}$, $W_{-1}^a$ and $W_{-2}^a$ in the case $k=0$, because when applied to some element of $W$, the result need not be in $C_1(W)$, but each of these modes strictly increases the generalized weight. As there is certainly no ``upper-truncation condition'' for the module $W$, the appearance of these modes in (\ref{sternie}) makes it seem possible that the complement of $C_1(W)$ in $W$ is infinite-dimensional. 

But fortunately, in this situation a theorem due to Buhl (see \cite{b0111}, Theorem~1) applies, which is a generalization of an earlier result of Gaberdiel and Neitzke \cite{gn0009}. It states the following (among other things): If a \VOA\ $V$ is $C_2$-cofinite, i.e.~$\text{dim}(V/C_2(V))<\infty$ with $C_2(V)=\text{span}\{u_{-2} v \,|\, u,v\in V\}$, then every weak $V$-module $W$ is spanned by elements of the form
\be\label{genx}
x_{-n_1}^1 \ldots x_{-n_k}^k w
\ee
with $w\in W$, $n_1 \geq \ldots \geq n_k > -L$, where $L$ is some fixed number, and the vectors $x^1, \ldots, x^k \in V$ are representatives of the elements of a basis of $V/C_2(V)$. In addition, if $n_j \leq 0$, then $n_i=n_j$ for at most $Q$ indices $i$, where $Q$ is another \textit{fixed} number. This last feature is the most important one for the present situation as it implies that only a limited number of powers of $L_{-1}$, $W_{-1}^a$ and $W_{-2}^a$ has to be considered if $V$ is $C_2$-cofinite. 

(Note that in (\ref{genx}) the convention for the indices of modes is the one used most often in the mathematics literature, i.e.~any vertex operator is expanded into a series $\sum_{n\in\Z}v_n x^{-n-1}$ regardless of the weight of the associated vector $v$. On the other hand, in the physics literature it is common to expand a field that is associated to a vector $u$ of weight $h$ into a series $\sum_{n\in\Z}u_n^{\text{{\tiny phys}}} x^{-n-h}$. The latter convention is used here only in the context of $\mathcal{W}$-algebras. When comparing results expressed in differing notations, the relation $u_n = u_{n-h+1}^{\text{{\tiny phys}}}$ is used.) 

The triplet algebra at $c=-2$ has the virtue of being $C_2$-cofinite because of the existence of certain singular vectors. Several authors (see \cite{gn0009} and \cite{miy0309}) have been aware of this fact for some time, and it was recently proven by Abe in \cite{abe0503}. The proof given here uses a different method. Indeed, the explicit form of six singular vectors at level 6 is known \cite{gk9606}, \cite{r9611}: 
\begin{align}\label{nullvector-2}
N^{ab} = & W_{-3}^a W_{-3}^b \Omega - \delta_{ab} \left( \frac{8}{9}\, L_{-2}^3 + \frac{19}{36}\, L_{-3}^2 + \frac{14}{9}\, L_{-4} L_{-2} - \frac{16}{9}\, L_{-6}\right) \Omega \nonumber \\
& \qquad + \I \varepsilon_{abc} \left( -2 W_{-4}^c L_{-2} + \frac{5}{4}\, W_{-6}^c \right) \Omega \; . 
\end{align}
In order to prove that $V$ is really $C_2$-cofinite, we first observe that in the above expression for the singular vector $N^{ab}$, because of (\ref{CminCn}) each term that it is made of is manifestly in $C_2(V)$ except for $W_{-3}^a W_{-3}^b \Omega$ and $L_{-2}^3 \Omega$. As any singular vector is divided out in the \VOA\ of interest, it follows that for $a\neq b$, 
$$
W_{-3}^a W_{-3}^b \Omega \in C_2(V) \quad \text{and} \quad \left( \big(W_{-3}^a\big)^2 - \big(W_{-3}^b\big)^2 \right) \Omega \in C_2(V) \; . 
$$
Since $W_{-3}^a$ leaves the space $C_2(V)$ invariant (recall	 (\ref{c1invariant})), $W_{-3}^a ((W_{-3}^a)^2 - (W_{-3}^b)^2 ) \Omega$ is an element of $C_2(V)$ as well. But this element can also be written as
\be\label{schwubbidubbi}
\left( W_{-3}^a \right)^3 \Omega - W_{-3}^a \left( W_{-3}^b  \right)^2 \Omega = \left( W_{-3}^a \right)^3 \Omega - W_{-3}^b W_{-3}^a W_{-3}^b \Omega + Y_{-6} W_{-3}^b \Omega \; , 
\ee
where $Y_{-6} = [W_{-3}^a, W_{-3}^b]$ applied to any vector $v\in V$ yields an element of $C_2(V)$ because in the commutator of modes of primary fields of weight 3 there can only appear modes corresponding to fields of weight less than or equal to 5 (see equation (\ref{WKommutator}) above). So in particular, the last term in (\ref{schwubbidubbi}) is in $C_2(V)$. In addition, the second last term in this equation also is in $C_2(V)$ as $W_{-3}^b$ leaves this space invariant. Hence, it follows that $(W_{-3}^a)^3 \Omega \in C_2(V)$, and we have 
\be\label{WinC2V}
\left( W_{-3}^a \right)^m \Omega \in C_2(V) \quad \text{for all } m \geq 3 \; .
\ee
From this and the fact that $((W_{-3}^a)^2 - \frac{8}{9}\, L_{-2}^3)\Omega$ is in $C_2(V)$ it follows that $(W_{-3}^a)^2 L_{-2}^3 \Omega \in C_2(V)$. Now using the invariance of $C_2(V)$ under $L_{-2}$ and $W_{-3}^a$ one more time it is easy to see that
$$
\left( \!\left( W_{-3}^a \right)^2 - \frac{8}{9}\, L_{-2}^3 \right)^2 \Omega = \left( \!\left( W_{-3}^a \right)^4 + \frac{64}{81}\, L_{-2}^6 - \frac{8}{9} \left( W_{-3}^a \right)^2 L_{-2}^3 - \frac{8}{9} L_{-2}^3 \left( W_{-3}^a \right)^2  \right) \Omega
$$
is an element of $C_2(V)$. But from the above discussion it is also clear that each term on the right-hand side apart from $\frac{64}{81}\, L_{-2}^6 \Omega$ is in $C_2(V)$, and so it follows that $ L_{-2}^6 \Omega$ must be an element of $C_2(V)$ as well. 

As a consequence of the reasoning of the last paragraph, sufficiently large powers of $L_{-2}$ and $W_{-3}^a$ (6 or maybe less in the first case, 3 or maybe less in the latter) applied to any element in $V$ yield elements in $C_2(V)$. Thus it is proven that $C_2(V)$ is finite-codimensional. 

Now that it has been shown that the prerequisite of Buhl's theorem is satisfied for the triplet algebra at $c=-2$, it can be used since by definition any object in $\mathcal{C}$ is a weak module and the elements in (\ref{sternie}) are of the same form as those in (\ref{genx}). This means that if it can be argued that $\omega$ and $W^a$ are \textit{not} in $C_2(V)$ and can thus be taken to be representatives of elements in a basis for $V/C_2(V)$, there actually \textit{is} some sort of an ``upper-truncation condition'', but for the exponents of the modes $L_{-1}$, $W_{-1}^a$ and $W_{-2}^a$ in (\ref{sternie}). So for $k=0$, it follows that only finitely many elements of the form (\ref{sternie}) span the ``$(k=0)$-part'' of $W$. This is exactly the statement that $W$ is $C_1$-cofinite. 

It remains to be seen that $\omega$ and $W^a$ are not in $C_2(V)$. For the moment, consider the possibility that $\omega$ \textit{is} in $C_2(V)$. Then there must be $u, v\in V$ such that $u_{-2}v = \omega$. By comparing weights on both sides, we arrive at the condition $\text{wt}\, u + \text{wt}\, v + 1 = 2$. But since the \VOA\ $V$ under consideration is of the form $V = \coprod_{n\in\N} V_{(n)}$ with $V_{(0)} = \C\Omega$, this condition says that either $u$ or $v$ must be (a scalar multiple of) the vacuum (and the other one of weight 1). This is not possible for the conformal vector, leading to a contradiction. By a similar reasoning, one also sees that $W^a \notin C_2(V)$. 

Finally, it needs to be shown that the chosen category is closed with respect to the contragredient functor. By the definition of the graded dual $W' = \coprod_{n\in\Z} (W_{[n]})^*$ it is clear that it is lower-truncated. In order to establish that it is also finitely generated, choose a minimal generating set $\{w_1, \ldots, w_N\} \subset W\in\text{ob}\,\mathcal{C}$ from a basis $\bigcup_{n\in\Z}B_n$ of $W$, where $B_n$ is a basis of $W_{[n]}$ for all $n\in\Z$. Then all $w\in W$ are linear combinations of elements of the form 
$$
\mathcal{M}_{n_1} \ldots \mathcal{M}_{n_k} w_i \; , 
$$
where $\mathcal{M}$ denotes the same as in (\ref{sternie}). Let $w'_1, \ldots, w'_N$ be the elements of the dual basis in $W'$ such that $\langle w'_i , w_j\rangle = \delta_{ij}$. Because of this, all $w'\in W'$ that may give a nonvanishing matrix element with some $w\in W$ must be linear combinations of elements of the form $\mathcal{M}'_{n_k} \ldots \mathcal{M}'_{n_1} w'_i$. To see this, assume that there is an element $\tilde{w}'\notin \{w'_1, \ldots, w'_N\}$ in $W'$ such that $\{\tilde{w}',w'_1, \ldots, w'_N\}$ is a subset of a minimal set of generating vectors of $W'$. It follows that $\langle \tilde{w}' , w_i \rangle = 0$ for all $i\in\{1,\ldots,N\}$ and thus 
\begin{align*}
\left\langle \mathcal{M}'_{-m_1} \ldots \mathcal{M}'_{-m_k} \tilde{w}' , \mathcal{M}_{-n_1} \ldots \mathcal{M}_{-n_l} w_i \right\rangle & = \left\langle \tilde{w}' , \mathcal{M}'_{m_k} \ldots \mathcal{M}'_{m_1} \mathcal{M}_{-n_1} \ldots \mathcal{M}_{-n_l} w_i \right\rangle \\
& = \delta_{\sum_i m_i, \sum_j n_j} \left\langle \tilde{w}' , \sum_{\{I \,|\, \text{wt}\, w_I = \text{wt}\, w_i\}} a_I w_I \right\rangle \\
& = 0 \; , 
\end{align*}
where $a_I\in\C$ are the coefficients that result from applying the commutation relations of the $\mathcal{M}$-modes. This means that the subspace of generalized weight $\text{wt}\,\tilde{w}'$ has a dimension that is strictly larger than the dimension of the corresponding subspace in $W$. But by the definition of the graded dual of $W$, these finite-dimensional subspaces must have the same dimension, so there cannot be an element $\tilde{w}'$ as above, and $\mathcal{C}$ is closed under the contragredient functor.

\subsection*{The triplet algebras at $\boldsymbol{c_{p,1}}$}

The triplet algebra at $c=-2$ is only the first member of an infinite family of triplet $\mathcal{W}$-algebras $\{\mathcal{W}(2,(2p-1)^{\times 3})\}_{p\geq 2}$ with central charge $c_{p,1} = 1-6(p-1)^2/p$, where for each $p\in\Z_{\geq 2}$ the three primary fields of weight $2p-1$ are a triplet under the action of the group $SO(3)$, which means that the structure constant $C_{W^a,W^b}^{W^c}$ is proportional to $\varepsilon_{abc}$ \cite{kau91}. It is the goal of this section to show that the above conditions (i), (ii) and (iii) are also satisfied in this general case. 

If one defines the category $\mathcal{C}$ analogously to the special case of $p=2$, one immediately sees that the conditions of quasi-finite dimensionality and of finitely generated lower-truncated generalized modules in $\text{ob}\,\mathcal{C}$ hold in the same way as before with the obvious generalization of the arguments. What requires additional work is to establish the $C_1$-cofiniteness of all objects in $\mathcal{C}$. 

Let $V_{\Delta}$ denote the \VOA\ associated to the $\mathcal{W}$-algebra $\mathcal{W}(2,\Delta^{\times 3})$ for a fixed $\Delta := 2p-1$ with $p\in\Z_{\geq 3}$. If $V_{\Delta}$ is $C_2$-cofinite, one can apply Buhl's theorem as in the case $p=2$, and any $V_{\Delta}$-module under consideration would be $C_1$-cofinite, which together with the other properties of $V_{\Delta}$ gives the existence and associativity of the nonmeromorphic \OPE. Compared with the case $p=2$, the difficulty of proving the $C_2$-cofiniteness of $V_{\Delta}$ stems from the lack of explicit expressions for singular vectors that are crucial for a proof of $C_2$-cofiniteness. A priori, it is not even clear whether such singular vectors at all exist for arbitrary $p\in\Z_{\geq 3}$. 

As it turns out, one can argue for the existence of certain singular vectors of weight $2(2p-1)$ with the help of the explicitly known character of $V_{\Delta}$ that was obtained in \cite{flohr9509}. By analyzing this character in detail one can show that for arbitrary $p\in\Z_{\geq 2}$, singular vectors of the form
\begin{align}\label{singularvectoratp}
N^{ab} & = W_{-\Delta}^a W_{-\Delta}^b \Omega + \delta_{ab}\big( \text{Virasoro-polynomial} \big) \Omega \nonumber \\
& \qquad + \varepsilon_{abc} \big( \text{Virasoro-}W_m^c \text{-polynomial} \big) \Omega
\end{align}
exist, where in the last term only summands with exactly one $W^c$-mode can appear (of course, equation (\ref{nullvector-2}) is of this form, too). 

In order to show that singular vectors as in (\ref{singularvectoratp}) really exist, we first recall the character
\be\label{characterVDelta}
\chi_{V_\Delta} (q) = \frac{q^{-1/24}}{\varphi(q)}\, \sum_{n\in\Z} (2n+1) q^{(2np+p-1)^2/(4p)}  
\ee
from \cite{flohr9509}. If we expand both this character and the vacuum Verma module character $\chi_{V_\Delta}^{\text{{\tiny Verma}}}(q)$ given by (\ref{vacVermachar}) into a formal power series in $q$ and compare the coefficients of $q^{(2p-1)+3}$ (times $q^{-c_{p,1}/24}$), we see that the dimensions of the homogeneous subspaces of weight $(2p-1)+3$ of the vacuum Verma module and the $\mathcal{W}$-algebra itself differ by 3. The reason for this is the following: From the Kac determinant it follows that the Virasoro algebra of central charge $c_{p,1}$ has an infinite set of highest weight modules where the highest weights are given by $h_{2k-1,1} = (k-1)(kp-1)$, $k\in\Z_+$. By a standard argument it follows that these modules have singular vectors at level $2k-1$. In particular, for $k=2$ the highest weight vectors of weight $2p-1$ can be identified with the vectors $W^a_{-\Delta} \Omega$ as we have $\Delta = 2p-1$. So because of the additional structure of the $\mathcal{W}$-algebra with its fields $W^a$, pure Virasoro modules are embedded into the full \VOA\ $\mathcal{W}(2,\Delta^{\times 3})$, and the difference of the dimensions above is due to the three singular vectors of weight $(2p-1)+3$. 

If these three vectors are divided out of the vacuum Verma module, we obtain a structure to which the character
\be\label{chartilde}
\tilde{\chi} (q) = q^{-c_{p,1}/24} \left( \frac{1}{\varphi_2(q)} + \frac{3 q^{2p-1} (1-q^3)}{\varphi(q) (\varphi_{2p-1}(q))^2} \right)
\ee
pertains, where we use the notation introduced in (\ref{truncvarphi}). The first term in this expression accounts for the action of the Virasoro algebra on the vacuum alone. The second term reflects the fact that beginning at level $2p-1$, the modes associated to the three distinct $W^a$-fields act nontrivially on the vacuum. With respect to the Virasoro algebra, this is a highest weight vector, which explains the factor $q^{2p-1}/\varphi(q)$. Furthermore, the factor $1-q^3$ is due to the singular vectors of weight $(2p-1)+3$ discussed above, and the term $(\varphi_{2p-1}(q))^{-2}$ comes from the action of the $W^a$-modes on the vacuum. The second power (and not the third) has to be taken here in order not to doubly count the contribution from the $W^a$-modes because of the three-fold multiplicity. 

Partially expanding both (\ref{characterVDelta}) and (\ref{chartilde}) into a formal power series yields
\begin{align*}
\chi_{V_\Delta} (q) & = \frac{q^{-c_{p,1}/24}}{\varphi(q)}\, \left( 1 - q + 3q^{2p-1} - 3q^{2p+2} + \mathcal{O}(q^{6p-2}) \right) \; , \\
\tilde{\chi} (q) & = \frac{q^{-c_{p,1}/24}}{\varphi(q)}\, \left( 1 - q + 3q^{2p-1} - 3q^{2p+2} + 6 q^{4p-2} + \mathcal{O}(q^{4p-1}) \right) \; . 
\end{align*}
We are interested in the dimensions of the homogeneous subspaces of weight $2\Delta = 4p-2$ described by these characters. Comparing the coefficients of $q^{4p-2}$ (times $q^{-c_{p,1}/24}$) by taking the relevant contributions from $(\varphi(q))^{-1} = \sum_{n\in\N}p(n)q^n$ into account, we see that these dimensions differ by 6. Thus we have found that six additional singular vectors of weight $2\Delta$ are divided out in $\mathcal{W}(2,\Delta^{\times 3})$. The reason that these vectors must involve a term with two $W^a$-modes is that there are no pure Virasoro singular vectors of weight $2\Delta$. Finally, the form of (\ref{singularvectoratp}) is a direct consequence of the $SO(3)$-structure of $\mathcal{W}(2,\Delta^{\times 3})$ \cite{kau91}. 

We now continue the proof of $C_2$-cofiniteness of the $\mathcal{W}$-algebras $\mathcal{W}(2,\Delta^{\times 3})$. As in the special case $p=2$ it is clear that nearly all possible vectors in the expression (\ref{singularvectoratp}) for the singular vector $N^{ab}$ are elements of $C_2(V_\Delta)$ because of the fact that $W_m^a \Omega = 0$ for all $m\geq -\Delta+1$. The only vectors for which this might not be true are $W_{-\Delta}^a W_{-\Delta}^b \Omega$ and $\alpha L_{-2}^{\Delta}\Omega$, the latter appearing in the $\delta_{aa}$-term in $N^{ab}$. If it can be shown that the coefficient $\alpha$ is not zero, the exact same reasoning as in the case $p=2$ can be applied to see that $V_\Delta$ is $C_2$-cofinite. So the question that remains to be answered is whether or not $\alpha\neq 0$. 

To find the correct answer, we first observe that the vertex operator to which a singular vector corresponds necessarily is a primary field. In particular, it is a quasi-primary field. As the vector $W_{-\Delta}^a W_{-\Delta}^b \Omega$ appears in the expression for the singular vector $N^{ab}$, the corresponding quasi-primary null-field must be a linear combination of quasi-primary fields, and one of these must be the normal-ordered product $\mathcal{N}(W^a,W^b)$. 

The next step is to note that the quasi-primary field $\mathcal{N}(W^a,W^b)$ alone cannot be the  null-field. To see this, we make use of the fact that the mode $L_1$ annihilates the vector $N^{ab}$. Thus, by expanding the null-field into modes, 
\begin{align}\label{L-1Naa1}
L_1 N^{aa} & = L_1 \left( W_{-\Delta}^a W_{-\Delta}^a \Omega + \big( \text{Virasoro-polynomial} \big) \Omega \right) \nonumber \\
& \stackrel{{\scriptscriptstyle \Delta-1}}{=} L_1 \left( W_{-\Delta}^a W_{-\Delta}^a \Omega + \beta L_{-4}L_{-2}^{\Delta-2} \Omega + \gamma L_{-3}^2 L_{-2}^{\Delta-3} \Omega \right) \nonumber \\
& \stackrel{{\scriptscriptstyle \Delta-1}}{=} 0 \; . 
\end{align}
Here, the symbol $\stackrel{{\scriptscriptstyle \Delta-1}}{=}$ has been introduced, which means ``equal to, modulo vectors with less than $\Delta-1$ modes applied to the vacuum $\Omega$''. For example, 
$$
\beta L_{-4}L_{-2}^{\Delta-2} \Omega + \gamma L_{-3}^2 L_{-2}^{\Delta-3} \Omega \stackrel{{\scriptscriptstyle \Delta-1}}{=} \beta L_{-4}L_{-2}^{\Delta-2} \Omega + \gamma L_{-3}^2 L_{-2}^{\Delta-3} \Omega + \delta L_{-4}^2 L_{-2}^{\Delta-4} \Omega \; . 
$$
So far, the values of the constants $\beta$ and $\gamma$ are unknown. If the null-field were equal to $\mathcal{N}(W^a,W^b)$, the coefficients $\beta = \beta_{WW}$ and $\gamma = \gamma_{WW}$ could be computed from the above formula (\ref{qpNOP}) for quasi-primary normal-ordered products in terms of the structure constant $C_{W^a, W^a}^{\mathcal{N}(T^{\Delta-1})}$. In principle, this constant can be computed for each $p\in\Z_{\geq 3}$ separately, but neither are such computations carried out easily nor is it necessary to know the exact value of the constant; only the information that it is not zero is crucial. 
 
With this, a straight-forward calculation using (\ref{L-1Naa1}) shows that 
$$
L_2 \left( W_{-\Delta}^a W_{-\Delta}^a  + \beta_{WW} L_{-4}L_{-2}^{\Delta-2}  + \gamma_{WW} L_{-3}^2 L_{-2}^{\Delta-3}  \right) \Omega \stackrel{{\scriptscriptstyle \Delta-1}}{=}\!\!\!\!\!\!\!\!\!\!\ \Big/ \;\;0 \; .
$$
So the field $\mathcal{N}(W^a,W^b)$ is quasi-primary but not primary and can thus not be the null-field. Instead, other quasi-primary fields must be added to $\mathcal{N}(W^a,W^b)$ to get the null-field. Of all these fields, only those are of immediate interest that yield primarity of the null-field \textit{at length} $\Delta-1$, i.e.~$L_2 N^{aa} \stackrel{{\scriptscriptstyle \Delta-1}}{=} 0$. Define $\mathcal{X}$ to be the set of all quasi-primary fields of weight $2\Delta$ except $\mathcal{N}(T^\Delta)$ in whose mode expansion appear Virasoro-monomials up to degree $\Delta-1$; in particular, $L_{-4} L_{-2}^{\Delta-2}$ is such a monomial. For example, $\mathcal{N}(\partial^2 T, \mathcal{N}(T^{\Delta-2})) \in \mathcal{X}$. Then the singular vector associated to the null-field satisfies the identity
\begin{align*}
N^{aa} \stackrel{{\scriptscriptstyle \Delta-1}}{=} & \left( \left(\mathcal{N}(W^a,W^a)\right)_{-2 \Delta} + \alpha \left(\mathcal{N}(T^\Delta)\right)_{-2 \Delta} + \sum_{X\in \mathcal{X}} k_X X_{-2 \Delta} \right) \Omega \\
\stackrel{{\scriptscriptstyle \Delta-1}}{=} & \left( \vphantom{\sum_{X\in\mathcal{X}}} W_{-\Delta}^a W_{-\Delta}^a + \alpha L_{-2}^\Delta + \left(\beta_{T^\Delta} + \beta_{WW}\right) L_{-4} L_{-2}^{\Delta-2} \right. \\
& \qquad \left. + \left(\gamma_{T^\Delta} + \gamma_{WW}\right) L_{-3}^2 L_{-2}^{\Delta-3} + \sum_{X\in\mathcal{X}} \left(\beta_X L_{-4} L_{-2}^{\Delta-2} + \gamma_X L_{-3}^2 L_{-2}^{\Delta-3}\right) \right) \Omega \; .
\end{align*}
(Note that there are no vectors of length $\Delta-1$ in $L_2 L_{-3}^2 L_{-2}^{\Delta-3}\Omega$, so the $\gamma$-terms do not have to be considered when $L_2$ acts on $N^{aa}$.) 

Now the assumption is made that $\alpha = 0$. Then one can use the fact that $L_2 N^{aa} = 0$ to find an explicit expression for the parameter $B := \sum_{X\in\mathcal{X}} \beta_X$ in terms of the structure constant $C_{W^a, W^a}^{\mathcal{N}(T^{\Delta-1})}$. (Fields $\mathcal{F}$ of weight $2\Delta-1$ like $\mathcal{N}(\partial T, T^{\Delta-2})$ with one derivative term need not be taken into account since the structure constants $C_{W^a,W^a}^{\mathcal{F}}$ for such fields vanish, see \cite{bfknrv}.) To do this, we need to know in which exact way $\beta_{WW}$ is proportional to $C_{W^a, W^a}^{\mathcal{N}(T^{\Delta-1})}$, so that we have $\beta_{WW} = \beta_{WW}' C_{W^a, W^a}^{\mathcal{N}(T^{\Delta-1})}$ with $\beta_{WW}'$ a nonzero constant whose exact value can be calculated to be $\beta_{WW}' = -\frac{(2\Delta-1)(\Delta-1)}{2(4\Delta-3)}$ by equation (\ref{qpNOP}). With this notation we arrive at
\begin{align*}
0  = L_2 N^{aa} & \stackrel{{\scriptscriptstyle \Delta-1}}{=} L_2 \left( W_{-\Delta}^a W_{-\Delta}^a + \left(\beta_{WW}' C_{W^a, W^a}^{\mathcal{N}(T^{\Delta-1})} + B \right) L_{-4} L_{-2}^{\Delta-2} \right) \Omega \\
& \stackrel{{\scriptscriptstyle \Delta-1}}{=} \left( \left[ L_2 , W_{-\Delta}^a W_{-\Delta}^a \right] + 6 \left(\beta_{WW}' C_{W^a, W^a}^{\mathcal{N}(T^{\Delta-1})} + B \right) L_{-2}^{\Delta-1} \right) \Omega \\
& \stackrel{{\scriptscriptstyle \Delta-1}}{=} \left( \left[ L_2 , W_{-\Delta}^a \right] W_{-\Delta}^a + 6 \left(\beta_{WW}' C_{W^a, W^a}^{\mathcal{N}(T^{\Delta-1})} + B \right) L_{-2}^{\Delta-1} \right) \Omega \\
& \stackrel{{\scriptscriptstyle \Delta-1}}{=} \left( \left(2(\Delta - 1) + \Delta\right) W_{2-\Delta}^a W_{-\Delta}^a + 6 \left(\beta_{WW}'C_{W^a, W^a}^{\mathcal{N}(T^{\Delta-1})}  + B \right) L_{-2}^{\Delta-1} \right) \Omega \\
& \stackrel{{\scriptscriptstyle \Delta-1}}{=} \left( (3\Delta - 2) \left[ W_{2-\Delta}^a, W_{-\Delta}^a\right] + 6 \left(\beta_{WW}' C_{W^a, W^a}^{\mathcal{N}(T^{\Delta-1})} + B \right) L_{-2}^{\Delta-1} \right) \Omega \\
& \stackrel{{\scriptscriptstyle \Delta-1}}{=} \left( (3\Delta - 2) C_{W^a, W^a}^{\mathcal{N}(T^{\Delta-1})} + 6 \left(\beta_{WW}' C_{W^a, W^a}^{\mathcal{N}(T^{\Delta-1})} + B \right) \right) L_{-2}^{\Delta-1} \Omega \; ,
\end{align*}
where it has been used in the last line that $p_{\Delta,\Delta,2 \Delta-2}(2-\Delta,-\Delta) = 1$. The above equation holds if and only if
\be\label{valueofB1}
B = - \frac{6 \Delta^2 - 8 \Delta + 3}{6(4 \Delta - 3)}\, C_{W^a, W^a}^{\mathcal{N}(T^{\Delta-1})} \; .
\ee

The idea to prove that $\alpha \neq 0$ now is to find another way to explicitly compute the value of $B$ that does \textit{not} agree with the one given in (\ref{valueofB1}). But before this is done it should be noticed that the parameters $\beta = \beta_{WW} + B$ and $\gamma = \gamma_{WW} + \sum_{X\in\mathcal{X}} \gamma_X$ can be expressed completely in terms of the structure constant $C_{W^a, W^a}^{\mathcal{N}(T^{\Delta-1})}$ and $B$: $\beta_{WW}$ and $\gamma_{WW}$ can be calculated by equation (\ref{qpNOP}), and from the fact that each field $X$ in $\mathcal{X}$ is quasi-primary (which means $L_1 X_{-2\Delta}\Omega \stackrel{{\scriptscriptstyle \Delta-1}}{=} 0$ among other things) it follows that $\sum_{X\in\mathcal{X}} \gamma_X = - \frac{5}{8}B$. As a consequence we have
\begin{subequations}\label{betagamma}
\begin{align}
\beta & = - \frac{1}{2}\, \frac{2 \Delta -1}{4 \Delta - 3} \, C_{W^a, W^a}^{\mathcal{N}(T^{\Delta-1})} (\Delta - 1)  + B \; , \label{beta} \\
\gamma & = - \frac{1}{2}\, \frac{2 \Delta -1}{4 \Delta - 3} \, C_{W^a, W^a}^{\mathcal{N}(T^{\Delta-1})} \left( (\Delta - 2)^2 - \frac{1}{2} (\Delta-2)(\Delta-3) \right) - \frac{5}{8} B \; . \label{gamma}
\end{align}
\end{subequations}
These relations will be made use of without explicit mention in the following. 

The vector $N^{aa}_{-2\Delta}\Omega$ is already completely known at length $\Delta-1$ up to the structure constant $C_{W^a, W^a}^{\mathcal{N}(T^{\Delta-1})}$, and the same situation will now be achieved for the vector $N^{aa}_{-2\Delta-1}\Omega$ as an intermediate step. For this, the relation $[L_m, \phi_n] = ((h-1)m-n)\phi_{m+n}$ with $m\in\{\pm 1,0\}$ for a quasi-primary field $\phi$ of weight $h$ is employed: at length $\Delta-1$ we see that $[L_{-1}, N^{aa}_{-2\Delta}]\Omega = L_{-1}N^{aa}$ is equal to
\begin{align}\label{L-1LHS}
& L_{-1} \left( W_{-\Delta}^a W_{-\Delta}^a + \beta L_{-4} L_{-2}^{\Delta-2} + \gamma L_{-3}^2 L_{-2}^{\Delta-3} \right) \Omega \nonumber \\
\stackrel{{\scriptscriptstyle \Delta-1}}{=} & \left( W_{-\Delta}^a \left[ L_{-1} , W_{-\Delta}^a \right] + \left[ L_{-1} , W_{-\Delta}^a \right] W_{-\Delta}^a + 3 \beta L_{-5} L_{-2}^{\Delta-2}  \right.  \nonumber \\
& \left. \qquad + (\Delta-2) \beta L_{-4} L_{-3} L_{-2}^{\Delta-3} + 4 \gamma L_{-4} L_{-3} L_{-2}^{\Delta-3} + (\Delta-3) \gamma L_{-3}^3 L_{-2}^{\Delta-4} \right) \Omega  \nonumber \\
\stackrel{{\scriptscriptstyle \Delta-1}}{=} & \left( \vphantom{L_{-2}^{\Delta-4}} 2 W_{-\Delta-1}^a W_{-\Delta}^a + \left[ W_{-\Delta}^a , W_{-\Delta-1}^a \right] \right.  \nonumber \\
& \left. \qquad + 3 \beta L_{-5} L_{-2}^{\Delta-2} + \left( (\Delta-2) \beta + 4 \gamma \right) L_{-4} L_{-3} L_{-2}^{\Delta-3} + (\Delta-3) \gamma L_{-3}^3 L_{-2}^{\Delta-4} \right) \Omega  \nonumber \\
\stackrel{{\scriptscriptstyle \Delta-1}}{=} & \,C_{W^a, W^a}^{\mathcal{N}(T^{\Delta-1})} p_{\Delta,\Delta,2\Delta-2}(-\Delta,-\Delta-1) \left( \vphantom{\binom{\Delta-1}{3}} (\Delta-1) L_{-5} L_{-2}^{\Delta-2} \right.  \nonumber \\
& \left. \qquad + (\Delta-1)(\Delta-2) L_{-4} L_{-3} L_{-2}^{\Delta-3} + \binom{\Delta-1}{3} L_{-3}^3 L_{-2}^{\Delta-4} \right) \Omega  \nonumber \\
& \qquad + \left( 3 \beta L_{-5} L_{-2}^{\Delta-2} + \left( (\Delta-2) \beta + 4 \gamma \right) L_{-4} L_{-3} L_{-2}^{\Delta-3} \right) \Omega \nonumber \\
& \qquad  + (\Delta-3) \gamma L_{-3}^3 L_{-2}^{\Delta-4} \Omega \; . 
\end{align}
But because of the quasi-primarity of the vector $N^{aa}$, this must also be equal to $N^{aa}_{-2\Delta-1}\Omega$. Of course the latter is not known explicitly, but at length $\Delta-1$ the relevant parameters can be inferred. Firstly, there is a contribution to $N^{aa}_{-2\Delta-1}\Omega$ from $\mathcal{N}(W^a,W^a)_{-2\Delta-1}\Omega$, and only the terms of length $\Delta-1$ will be of importance here. Secondly, the contribution of the fields in $\mathcal{X}$ has to be taken into account. Computing this contribution exactly would require the knowledge of the exact values of the parameters $k_X$ in 
$$
N^{aa} \stackrel{{\scriptscriptstyle \Delta-1}}{=} \mathcal{N}(W^a,W^a)_{-2\Delta} \Omega + \sum_{X\in\mathcal{X}} k_X X_{-2\Delta} \Omega \; .  
$$
These are not available, but all we really need to know in this case are the coefficients of the relevant monomials at length $\Delta-1$. Denoting these coefficients by $\xi_i$, $i\in\{1,2,3\}$, (\ref{L-1LHS}) is also equal to
\begin{align}\label{L-1RHS}
& \mathcal{N}(W^a,W^a)_{-2\Delta-1}\Omega + \left( \xi_1 L_{-5} L_{-2}^{\Delta-2} + \xi_2 L_{-4} L_{-3} L_{-2}^{\Delta-3} + \xi_3 L_{-3}^3 L_{-2}^{\Delta-4} \right) \Omega \nonumber \\
\stackrel{{\scriptscriptstyle \Delta-1}}{=} & - \frac{1}{4}\, C_{W^a, W^a}^{\mathcal{N}(T^{\Delta-1})} \frac{2\Delta-1}{4\Delta-3} (-n-2\Delta+2)(-n-2\Delta+1)\Big|_{n=-2\Delta-1} \nonumber \\
& \qquad \cdot \left( (\Delta-1) L_{-5} L_{-2}^{\Delta-2} + (\Delta-1)(\Delta-2) L_{-4} L_{-3} L_{-2}^{\Delta-3} \vphantom{+ \binom{\Delta-1}{3} L_{-3}^3 L_{-2}^{\Delta-4} \Omega}\right. \nonumber \\
& \left. \qquad\qquad + \binom{\Delta-1}{3} L_{-3}^3 L_{-2}^{\Delta-4} \right) \Omega \nonumber \\
& \qquad + \left( \xi_1 L_{-5} L_{-2}^{\Delta-2} + \xi_2 L_{-4} L_{-3} L_{-2}^{\Delta-3} + \xi_3 L_{-3}^3 L_{-2}^{\Delta-4} \right) \Omega \; .
\end{align}
Now comparing the coefficients of the vectors $L_{-5} L_{-2}^{\Delta-2}\Omega$, $L_{-4} L_{-3} L_{-2}^{\Delta-3}\Omega$ and $L_{-3}^3 L_{-2}^{\Delta-4}\Omega$ in (\ref{L-1LHS}) and (\ref{L-1RHS}) yields 
\begin{subequations}\label{xi}
\begin{align}
\xi_1 & = \frac{1}{2} \left( 6 B + C_{W^a, W^a}^{\mathcal{N}(T^{\Delta-1})} (\Delta-1) \right)  \; , \label{xi1} \\
\xi_2 & = \frac{1}{2} \left( - 9 B + 2 B \Delta + C_{W^a, W^a}^{\mathcal{N}(T^{\Delta-1})} \left(\Delta^2 - 3 \Delta + 2 \right) \right)  \; , \label{xi2} \\
\xi_3 & = \frac{1}{24} \left(45 B - 15 B \Delta + C_{W^a, W^a}^{\mathcal{N}(T^{\Delta-1})} \left( 2 \Delta^3 - 12 \Delta^2 + 22 \Delta - 12 \right)  \right)  \; . 
\end{align}
\end{subequations}
With this knowledge of both vectors $N^{aa}_{-2\Delta}\Omega$ and $N^{aa}_{-2\Delta-1}\Omega$ at length $\Delta-1$, now one last piece of information can be utilized in order to find another way to compute $B$. Until now, only the quasi-primarity of the null-field has been used. But actually it is also primary, i.e.~the relation $[L_m, N^{aa}_n] = ((2\Delta-1)m-n)N^{aa}_{m+n}$ holds for all $m,n\in\Z$. In particular, this is true for $m=2$ and $n=-2\Delta-1$, and thus 
\begin{align}\label{L-2Null}
0 & = (6\Delta-1) N^{aa}_{-2\Delta+1} \Omega = \left[L_2 , N^{aa}_{-2\Delta-1} \right] \Omega = L_2 N^{aa}_{-2\Delta-1} \Omega \nonumber \\
& \stackrel{{\scriptscriptstyle \Delta-1}}{=} - \frac{3}{2} C_{W^a, W^a}^{\mathcal{N}(T^{\Delta-1})} \frac{2\Delta-1}{4\Delta-3} \left( 7 (\Delta-1) + 6 (\Delta-1)(\Delta-2) \right) L_{-3} L_{-2}^{\Delta-2} \Omega \nonumber \\
& \qquad\quad\ + \left( 7 \xi_1 + 6 \xi_2 \right) L_{-3} L_{-2}^{\Delta-2} \Omega + \left[ L_2 , N^{(\Delta)}(W^a,W^a)_{-2\Delta-1} \right] \Omega \nonumber \\
& \stackrel{{\scriptscriptstyle \Delta-1}}{=} - \frac{3}{2} C_{W^a, W^a}^{\mathcal{N}(T^{\Delta-1})} \frac{2\Delta-1}{4\Delta-3} \left( 7 (\Delta-1) + 6 (\Delta-1)(\Delta-2) \right) L_{-3} L_{-2}^{\Delta-2} \Omega \nonumber \\
& \qquad\quad\ + \left( 7 \xi_1 + 6 \xi_2 \right) L_{-3} L_{-2}^{\Delta-2} \Omega + \left[ L_2 , W^a_{-\Delta} W^a_{-\Delta-1} + W^a_{-\Delta-1} W^a_{-\Delta} \right] \Omega \nonumber \\
& \stackrel{{\scriptscriptstyle \Delta-1}}{=} - \frac{3}{2} C_{W^a, W^a}^{\mathcal{N}(T^{\Delta-1})} \frac{2\Delta-1}{4\Delta-3} \left( 7 (\Delta-1) + 6 (\Delta-1)(\Delta-2) \right) L_{-3} L_{-2}^{\Delta-2} \Omega \nonumber \\
& \qquad\quad\ + \left( 7 \xi_1 + 6 \xi_2 \right) L_{-3} L_{-2}^{\Delta-2} \Omega \nonumber \\
& \qquad\quad\ + \left( \left[ L_2 , W^a_{-\Delta} \right] W^a_{-\Delta-1} + \left[ L_2, W^a_{-\Delta-1} \right] W^a_{-\Delta} \right) \Omega \nonumber \\
& \stackrel{{\scriptscriptstyle \Delta-1}}{=} - \frac{3}{2} C_{W^a, W^a}^{\mathcal{N}(T^{\Delta-1})} \frac{2\Delta-1}{4\Delta-3} \left( 7 (\Delta-1) + 6 (\Delta-1)(\Delta-2) \right) L_{-3} L_{-2}^{\Delta-2} \Omega \nonumber \\
& \qquad\quad\ + \left( 7 \xi_1 + 6 \xi_2 \right) L_{-3} L_{-2}^{\Delta-2} \Omega + \left( 2(\Delta-1) + \Delta \right) \left[ W^a_{-\Delta+2} , W^a_{-\Delta-1} \right]  \Omega \nonumber \\ 
& \qquad\quad\ + \left( 2(\Delta-1) + \Delta + 1 \right) \left[ W^a_{-\Delta+1} , W^a_{-\Delta} \right]  \Omega \nonumber \\
& \stackrel{{\scriptscriptstyle \Delta-1}}{=} - \frac{3}{2} C_{W^a, W^a}^{\mathcal{N}(T^{\Delta-1})} \frac{2\Delta-1}{4\Delta-3} \left( 7 (\Delta-1) + 6 (\Delta-1)(\Delta-2) \right) L_{-3} L_{-2}^{\Delta-2} \Omega \nonumber \\
& \qquad\quad\ + \left( 7 \xi_1 + 6 \xi_2 \right) L_{-3} L_{-2}^{\Delta-2} \Omega \nonumber \\
& \qquad\quad\ + (3\Delta-2) C_{W^a, W^a}^{\mathcal{N}(T^{\Delta-1})} p_{\Delta,\Delta,2\Delta-2}(2-\Delta,-\Delta-1) (\Delta-1) L_{-3} L_{-2}^{\Delta-2} \Omega \nonumber \\
& \qquad\quad\ + (3\Delta-1) C_{W^a, W^a}^{\mathcal{N}(T^{\Delta-1})} p_{\Delta,\Delta,2\Delta-2}(1-\Delta,-\Delta) (\Delta-1) L_{-3} L_{-2}^{\Delta-2} \Omega \; ,
\end{align}
where in this case the term $N^{(\Delta)}(W^a,W^a)_{-2\Delta-1}\Omega$ (recall equation (\ref{NOPModen})) \textit{does} lead to a contribution at length $\Delta-1$, in contrast to the situation in equation (\ref{L-1RHS}). Now using (\ref{betagamma}) and (\ref{xi}) in (\ref{L-2Null}) yields the following alternate expression for the parameter $B$: 
$$
B = - \frac{12 \Delta^2 - 18 \Delta + 7}{4( 4 \Delta- 3)} \, C_{W^a, W^a}^{\mathcal{N}(T^{\Delta-1})} \; .
$$
This can only be in agreement with (\ref{valueofB1}) for $C_{W^a, W^a}^{\mathcal{N}(T^{\Delta-1})} = 0$, which is not the case. Thus, the assumption $\alpha=0$ leads to a contradiction and $V_\Delta$ is $C_2$-cofinite. 

\vspace{0.2cm}

We summarize our results in the following theorem. 

\vspace{0.2cm}

\noindent\textit{Theorem}. For all $p\in\Z_{\geq 2}$, the nonmeromorphic \OPE\ exists and is associative for the \VOA\ $\mathcal{W}(2,(2p-1)^{\times 3})$. Furthermore, all these \VOA s are $C_2$-cofinite.

\end{document}